\documentclass[usenatbib]{mnras}

\usepackage{newtxtext,newtxmath}
\usepackage{xspace}
\usepackage{bm}

\usepackage{float}
\usepackage[T1]{fontenc}
\usepackage{amsmath}
\usepackage{hyperref}

\DeclareRobustCommand{\VAN}[3]{#2}
\let\VANthebibliography\thebibliography
\def\thebibliography{\DeclareRobustCommand{\VAN}[3]{##3}\VANthebibliography}

\newcommand{\LODI}{\texttt{LODI}}
\newcommand{\HALO}{\texttt{HALOgen}}
\newcommand{\mpc}{~\text{Mpc $h$}^{-1}}

\usepackage{graphicx}	
\usepackage{amsmath}	
\usepackage{algorithm}
\usepackage{algpseudocode}

\title[\LODI]{Large, fast and accurate HI intensity maps with latent overlap diffusion}

\author[Mishra, Trotta \& Viel]{
Satvik Mishra,$^{1}$\thanks{E-mail: samishr@sissa.it}
Roberto Trotta,$^{1,2,3,4}$
Matteo Viel$^{5,1,2,3,6,7}$
\\
$^{1}$Theoretical and Scientific Data Science, SISSA, Via Bonomea 265, 34136 Trieste, Italy\\
$^{2}$INFN -- National Institute for Nuclear Physics, Via Valerio 2, 34127 Trieste, Italy\\
$^{3}$ICSC - Centro Nazionale di Ricerca in High Performance Computing, Big Data e Quantum Computing, Via Magnanelli 2, Bologna, Italy\\
$^{4}$Astrophysics Group, Physics Department, Blackett Lab, Imperial College London, Prince Consort Road, London SW7 2AZ, UK\\
$^{5}$ Astroparticle and Gravitational Physics Group, SISSA, Via Bonomea 265, 34136 Trieste, Italy\\
$^{6}$INAF -- Osservatorio Astronomico di Trieste, Via G. B. Tiepolo 11, I-34143 Trieste, Italy\\
$^{7}$IFPU -- Institute for Fundamental Physics of the Universe, Via Beirut 2, I-34151 Trieste, Italy\\
}

\pubyear{\the\year{}}

\begin{document}
\label{firstpage}
\pagerange{\pageref{firstpage}--\pageref{lastpage}}
\maketitle

\begin{abstract}
The distribution of 21 cm emission from neutral hydrogen is a powerful cosmological and astrophysical probe, as it traces the underlying dark matter and cold gas distributions throughout cosmic times. However, the prediction of observable signals is hindered by the large computational costs of the required hydrodynamic simulations. We introduce a novel machine learning pipeline that, once trained on a hydrodynamical simulation, is able to generate both halo mass density maps and the three-dimensional 21 cm brightness temperature signal, starting from a dark matter-only simulation. We use an attention-based ResUNet (\HALO) to predict dark matter halo maps, which are then processed through a trained conditional variational diffusion model (\LODI) to produce 21 cm brightness temperature maps. \LODI\ is trained on smaller sub-volumes that are then seamlessly combined in 512-times larger volume using a new method, called `latent overlap'. We demonstrate that, once trained on 25$^3$ $\left(\text{Mpc}/h\right)^3$ volume simulations, we are able to predict the 21 cm power spectrum on an unseen dark matter map (with the same cosmology) to within $10\%$ for wavenumbers $k\leq10~\text{$h$~Mpc}^{-1}$, deep inside the non-linear regime, with a computational effort of the order of two minutes. While demonstrated on this specific volume, our approach is designed to be scalable to arbitrarily large simulations.
\end{abstract}




\section{Introduction}
Neutral (atomic) hydrogen (HI) plays an important role in cosmology and structure formation processes: its distribution follows the underlying matter density field, and for most of the cosmic history it constitutes the reservoir of baryons to fuel star formation. This makes it a powerful and novel tracer of the large-scale structure (LSS) of the Universe~\citep[e.g.][]{review,Ansari_2012,Santos:2015,villa14}. While Cosmic Microwave Background (CMB) observations~\citep{WMAP,planck:2018,ACT} and galaxy redshift surveys~\citep[e.g.][]{BOSS:2016} have significantly constrained the parameters of the standard $\Lambda$CDM cosmological model, key questions are still unresolved. In particular, the fundamental nature of dark matter and dark energy is still unknown, and persistent tensions between different cosmological measurements have yet to be fully understood, including the recent tentative evidence for evolving dark energy ~\citep[e.g.][]{riess:2019,Verde:2019, wong:2019,DESIde}. Mapping the large-scale distribution of HI and tracking its evolution over cosmic time offer a complementary approach to traditional galaxy surveys, by probing large volumes at high redshifts, and thus potentially providing stress tests for many cosmological models in new regimes ~\citep[e.g.][]{bull:2015,villa15,obuljen_high-redshift_2018, berti_constraining_2022,Villaescusa-Navarro:2016}.

The 21 cm line, which results from the spin-flip transition within the hyperfine structure of the ground state of neutral hydrogen (see~\citet{Furlanetto:2006}), is redshifted by the cosmological expansion and can be observed at radio wavelengths. Many experiments, including interferometric arrays like CHIME~\citep{Bandura:2014gwa,CHIMEdetection}, CHORD, and HIRAX~\citep{Newburgh:2016mwi}, and single-dish instruments such as GBT~\citep{Masui2013,Wolz2022} and FAST~\citep{Hu:2019okh}, aim to detect this signal using intensity mapping (IM) techniques~\citep{Bharadwaj:2000,Battye:2004,McQuinn:2005,Chang:2007,Seo:2009fq,Battye:2013,Kovetz:2017}. Several of these efforts have already achieved detections through cross-correlation with optical galaxy surveys~\citep{Chang2010,Masui2013,Anderson2018,Wolz2022,Cunnington:2022,Paul2023}.
Radio cosmology is also a key science objective for the SKA Observatory (SKAO),\footnote{\url{https://www.skao.int/}} which will comprise two major arrays: SKA-Low in Australia and SKA-Mid in South Africa. In particular, SKA-Mid, when operated in single-dish mode~\citep[e.g.][]{Santos:2015,Bacon:2018}, will be capable of conducting 21 cm IM surveys across cosmologically relevant scales out to redshift $z \sim 3$. Currently under construction, the SKAO has a working precursor, MeerKAT, which is already contributing through its cosmological IM survey, MeerKLASS~\citep{Santos:2017}. 
Early results from MeerKAT data have been encouraging~\citep{Wang:2021,irfan2022}, including a first cross-correlation detection with WiggleZ galaxy data~\citep{Cunnington:2022uzo}. 

Alongside several technical efforts mainly focused on foreground modeling and mitigation, refining the forecast capabilities of 21 cm IM, both as a standalone probe and in combination with other cosmological observables, is crucial (e.g~\citep{carucci_cross-correlation_2017,berti_21cm_2023,berti24}). These forecasts are essential not only for strengthening the scientific case for 21 cm IM radio cosmology, but also in order to optimize the design and strategy of upcoming surveys.
The theoretical modeling of the 21 cm signal follows three different approaches: halo models (see~\citet{padmanabhan_hi_2023}), perturbation theory (e.g.~\citep{obuljen23}) or hydrodynamical simulations of structure formation incorporating the relevant physical processes (e.g.~\citep{villaescusa-navarro_ingredients_2018}). 
    Each of these methods has both advantages and disadvantages. For example, modeling the IM signal requires both large volumes and high resolution to fully resolve the physics of the small mass ($\sim$ 10$^{10}$ M$_{\odot}/h$) dark matter halos that host HI, something that is difficult to achieve with full hydrodynamical simulations due to the range of scales. {In the context of halo models, extensive work has been expended to study the mass limit for halos to retain HI, for example by looking at the damped Lyman-$\alpha$ systems statistics, both from the point of view of halo models~\citep{Padmanabhan_2016} and observationally~\citep{Dev_2023,Obuljen_2019}}. {In addition to these semi-analytical and perturbative frameworks, several forward-modeling approaches have been developed to predict the HI field directly from dark-matter simulations by
prescribing  or calibrating the $M_{\rm HI}(M)$ relation and other parameters on observations or hydrodynamical runs~\citep[e.g.][]{hitz2025fastsimulationcosmologicalneutral,li2022theoreticalmodelsatomichydrogen,spinelli_atomic_2020}. These methods however extend predictions to large volumes by applying calibrated prescriptions for the HI–halo connection, rather than learning the full field-level mapping from simulations. } 
Thus, a method capable of learning the small scale physics from state-of-the-art simulations while at the same time reaching large scales and volumes would be of great importance.

In the context of machine learning (ML) application, {progress in neural-based modeling has been recently achieved at large scales and for large volumes~\citep[e.g.][]{teachingdarkmatter,dreams,charm} .} In the context of 21 cm IM theoretical modeling, there have been several recent efforts aim at predicting the brightness temperature signature at the field level, including fast generative models of HI maps using normalizing flows \citep{hassan_hiflow_2022}, 
the neural approach of \citet{wadekar_hinet_2021}, and a generative adversarial framework introduced by \citet{andrianomena_emulating_2022}.
Diffusion models (e.g.~\citep{kingma_variational_2023}) have thus far found very limited applications~\citep{ono_debiasing_2024}, and none in the realm of 21 cm IM. In this work, we develop a novel, physically motivated ML pipeline, which in a first step uses a U-Net to learn the mapping between dark matter and halos; in a second step, it employs a custom-designed diffusion model to predict the 21 cm signal, down to very small, non-linear scale. Predicting both halos and 21 cm signal is crucial in the context of using such machinery for cross-correlation studies of the IM signal with other LSS tracers (like lensing, galaxies, etc.), which is the ultimate of aim of our work. {This work should therefore be viewed as the first demonstration of a diffusion-based, halo-conditioned approach to 21 cm IM, complementary to existing forward modeling methods.}

The paper is organized as follows: in Section \ref{methodology} we describe in the the methodology used, including the ML pipeline, U-Net architecture, diffusion model with the novel latent overlap method, simulations used, loss function, networks training and validation of the pipeline. in Section \ref{results} we present our results in terms of illustrative halo and HI maps, and quantitative comparison of power spectra predicted for held-out dark matter maps. We conclude in Section \ref{conclusions}, where we also indicate promising avenues for future work and applications of our method.

\section{Methodology}
\label{methodology}
In this section, we explain the machine learning technique, the mock datasets used, and the training methods implemented.

Our strategy is the following: starting from a dark matter only map, we use a U-Net architecture to predict the halo map, as described in section~\ref{sec:dm_to_halo}.

We then use the halo map as input to a variational diffusion model trained to inpaint the HI brightenss temperature, as described in section~\ref{sec:halo_to_HI}.

\begin{figure*}
    \centering
    \includegraphics[width=0.6\linewidth]{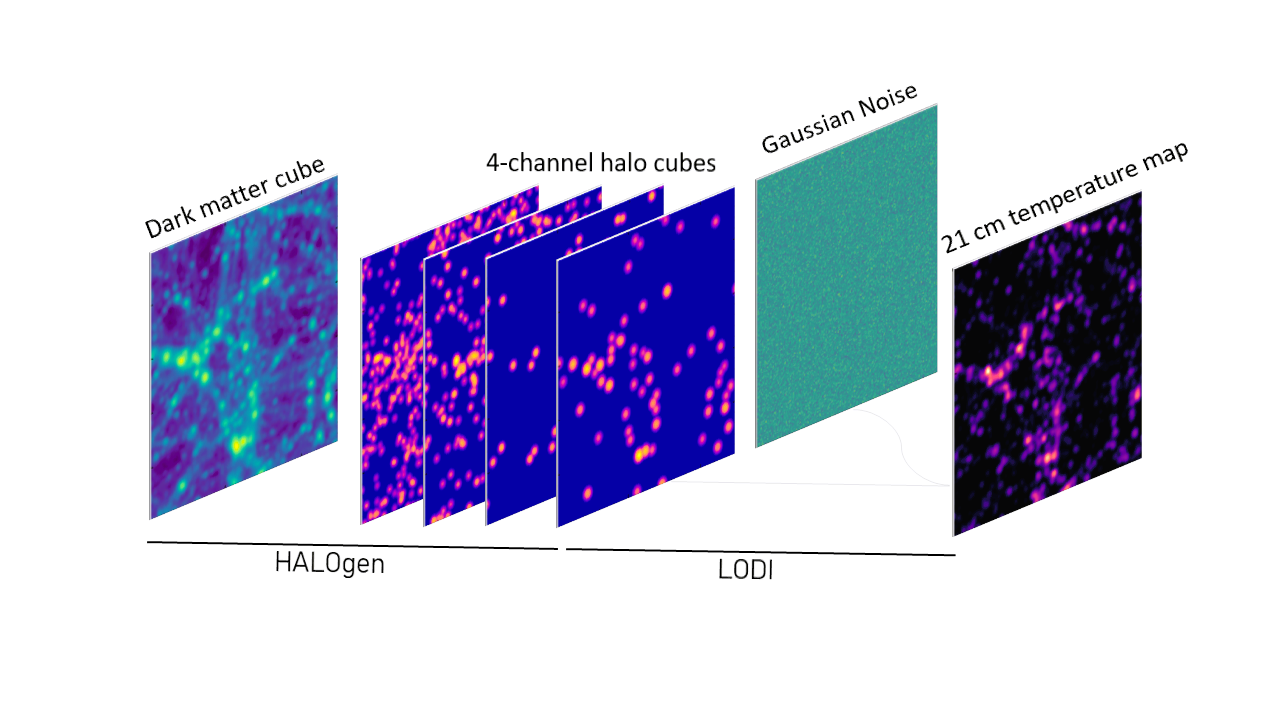}
    \caption{An overview of our generative pipeline, starting from dark matter particle distribution as produced by an N-body code to the final 21cm intensity map: in the first step, a ResNet with attention bottleneck (\HALO) is used to predict the dark matter halo mass density in four mass channels; subsequently, a variational diffusion model with latent overlap (\LODI) generates the 21 cm brightness temperature map.}
    \label{fig:pipeline}
\end{figure*}

An overview of the full pipeline is presented in Fig.~\ref{fig:pipeline}.

\subsection{\HALO: from dark matter to halo maps} \label{sec:dm_to_halo}

\subsubsection{U-Net architecture}

A U-Net, introduced by~\citet{ronneberger_u-net_2015} is a deep hierarchical convolutional neural network with encoder and decoder layers, used to perform image-to-image translation tasks -- an ideal architecture for our aim of predicting halo maps from $\boldsymbol{\rho}_\text{DM}$ maps. 

We use ResNet blocks~\citep{he_deep_2015} as the fundamental blocks for the U-Net architecture, as well as an attention block at the bottleneck motivated by~\citet{petit_u-net_2021}. The attention block enhances the ability of the model to capture and highlight complex structure and long-range dependencies within the data. 
\begin{figure*}
    \centering
    \includegraphics[width=0.6\linewidth]{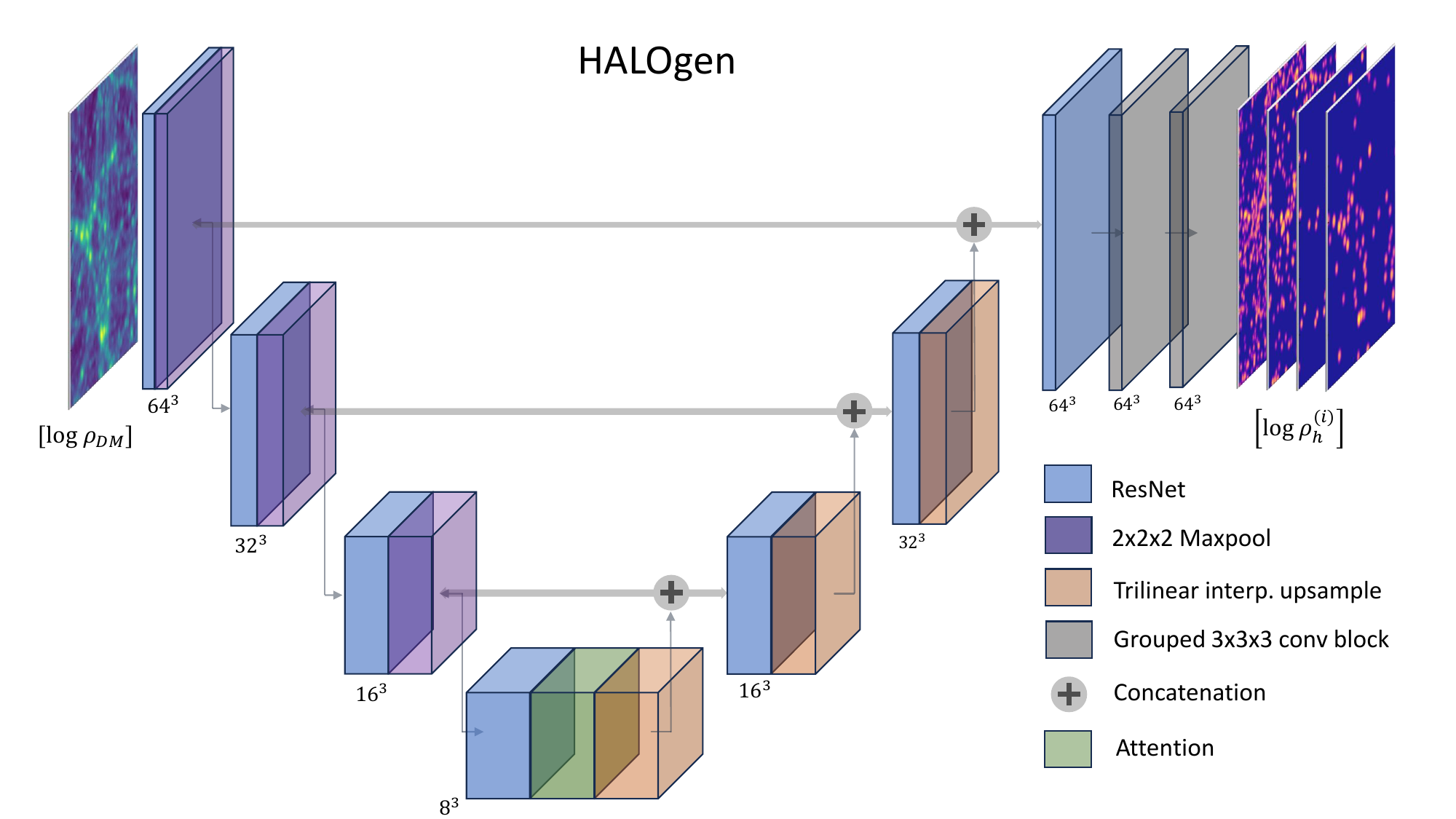}
    \caption{Overview of the \HALO\ (Halo Assignment and Generation with U-Net) architecture used for halo assignment from an input DM-only map, employing grouped 3D convolutions (gray blocks) and an attention mechanism at the bottleneck. Upsampling is done via trilinear interpolation to avoid artifacts in the upsampled map. Horizontal lines represent skip connections.{The input data has shape of $64^3$ voxels while the bottleneck has a spatial shape of $8^3$.} }
    \label{fig:Unethalo}
\end{figure*}
The architecture, depicted in Fig.~\ref{fig:Unethalo}, consists of 3 encoding blocks with grouped convolutions, which process a 3D $\boldsymbol{\rho}_\text{DM}$ input of size $64^3$ voxels. The input is progressively downsampled through three stages, reaching a bottleneck representation of size of $8^3$ voxels, with $2\times2\times2$ kernel max-pooling layers employed for downsampling to reduce spatial resolution while preserving dominant features. 
For the up-sampling leg of the architecture, we use trilinear interpolation instead of the more commonly used transposed convolutions to avoid checkered board artifacts -- grid-like patterns that can arise when upsampling images. Trilinear interpolation estimates voxel values by computing a weighted average of the eight nearest neighbors, ensuring a smoother reconstruction. To enhance high-level feature preservation, we incorporate skip connections (represented as horizontal grey lines in Fig.~\ref{fig:Unethalo}) after upsampling, whose aim is to transfer directly feature maps from corresponding encoder layers to decoder layers, thus bypassing the bottleneck. 

The last decoder block leads to a set of grouped convolution layers that transform the output into 4 channels, each corresponding to a halo map density field $\boldsymbol{\rho}_{\text{h}}^{(i)}$, where $i=1,\dots,4$ is the channel number. Each channel describes halos within a top-hat mass range (a bin). Attributing halos to different mass bins is necessary to reduce the dynamic range of the relationship between halo mass and HI mass, and to account for its strong dependence on halo mass: halos with mass $M_{\text{halo}}<3\times10^{10}~M_\odot $ contribute $\approx 95 \%$ of the total number count of halos, but their combined HI mass is only $\approx 5 \%$ of the total HI in the simulation. At the other end of the mass distribution, halos of mass $\geq 10^{12} M_\odot$ populate only about $\approx 0.5 \%$ of the simulation but account for more than $ 25 \%$ of the HI mass. Our binning both reduces the dynamic range spanned by the $M_{\text{HI}}$ to $M_{\text{halo}}$ relationship (thus facilitating learning) and also accounts for the halo-mass dependency of the relation.
{The HI mass function, as determined by~\citet{Ma_2025_FASHI,Jones_ALFA}, features a knee HI mass of $\approx7.5\times10^9M_\odot$, with the power-law slope for low mass being $\alpha=-1.3$, indicating that contribution from HI regions smaller than the knee HI mass is non-negligible. Following~\citet{villaescusa-navarro_ingredients_2018}, the HI cutoff mass for our case is $\approx10^8M_\odot$, which is almost two orders of magnitude smaller than the knee mass determined observationally. This provides reassurance that we are accounting for most of the HI regions.} 

The (fixed) boundaries of the top-hat mass ranges are chosen to obtain similar HI mass within each bin\footnote{This remains true, up to a few percent, for all maps in the cross-validation set of the simulation, which share the same astrophysical parameters. Generalizing this approach across different parameter values will be the focus of future work.},  and are given by:
\begin{equation}
\left[3 \times 10^{10}, 10^{11}, 5 \times 10^{11}, 10^{12}, \max \left( M_{\text{halo}} \right) \right] \ M_\odot \, . 
\label{eq:halo_bins}
\end{equation}
{where the maximum value of $M_\text{halo}$ is $\approx 1.7\times10^{14}M_\odot$.} 
We use group convolution and group normalization layers to enforce that output channels be independent; experimentation with mixing the information between different groups (4 or multiples of 4) has shown that this degrades performance. The output of the network has a voxel dimension of $4\times 64^3$. {Compared to larger simulations such as TNG300, which host halos up to $\approx 10^{15}M_\odot$, the smaller CAMELS volumes do not sample such large systems. While this restricts the direct modeling of the rarest, most massive clusters, CAMELS provides a more machine-learning–friendly baseline, and our approach can in the future be extended to larger simulations or large cluster regions. }

To create the full $256^3$ voxels map, we take  individual $64^3$ voxels sub-volume dark matter fields and feed them to our model. Then we use a stride of 16 pixels in every cartesian direction to pick our next $64^3$ box. The overlapping pixels are averaged over to get the final output. 

\subsubsection{Training dark matter and halo mass density maps}

To obtain the input DM mass density map, we take the positions for each DM particle from the DM-only simulations with side 25$h^{-1}$ Mpc and convert them to 3 dimensional $\boldsymbol{\rho}_\text{DM}$ fields of size $256^3$ voxels, by using the cloud in cell (CIC) algorithm. {The cloud-in-cell (CIC) algorithm assigns particle masses to a regular grid by distributing each particle’s mass to the eight nearest grid points, weighted linearly by the particle’s relative distance from each grid point. This produces a continuous density field while conserving total mass.} Our field resolution is thus $\approx 0.1 h^{-1}$ Mpc. We use for our model training the CAMELS dataset from~\citet{villaescusa-navarro_camels_2022}, a suite of hydrodynamical and N-body cosmological simulations. More detail related to the simulations we use is provided in section~\ref{dataset}.

To determine the halo mass density field $\rho_\text{h}$, we use the SUBFIND algorithm~\citep{springel_populating_2000}, available in the simulation suite to extract the halo center of mass positions and then use the same CIC algorithm, weighted by the mass of the halos. We manually create the four halo channels using the bins given in~Eq.~\eqref{eq:halo_bins}, and apply the CIC algorithm on each bin separately.

Maps are smoothed with a Gaussian kernel of fixed radius $R = 0.2 ~\text{Mpc $h$}^{-1}$ to increase the signal to noise ratio (SNR) and remove irrelevant small-scale features to help training.

\subsubsection{\HALO~loss function} 

The loss function needs to deal with the sparsity of the $\boldsymbol{\rho}_{\text{h}}^{(i)}$ maps. Furthermore, a given DM voxel may be associated with multiple halos of different masses (and falling into different halo mass bins), making it challenging for the model to disentangle their respective contributions. We call this phenomenon `channel mixing', i.e., the fact that halo information from one channel leaks into adjacent channels. Moreover, since the binning scheme is not explicitly encoded within the neural network architecture, the model must learn to distinguish halos across adjacent bins without misclassifications. 
Ensuring proper bin differentiation is crucial to maintaining the physical consistency of the model predictions. 

To address the above issues, we create a custom-weighted mask, $m_w^i$ ($i=1, \dots, 4$), associated to each halo channel $i$. The purpose of the mask is to give higher weight to pixels with halos in channel $i$, while enforcing a negligible contribution to the loss from halos in neighboring channels, thus helping both with sparsity and channel separation.

For a given halo channel $i$, with halo mass density field $\boldsymbol{\rho}_{\text{h}}^{(i)}$, we define its weight as:
\begin{equation} \label{eqn:mask}
    w^i = 
    \frac{N_\text{mesh}^3}{\displaystyle \sum_{j = \max(1,i-1)}^{\min(4,i+1)} \sum_{k=1}^{N_\text{mesh}} {\boldsymbol{\theta}}(\rho_h^{(j,k)} - T_c^{(j)})} \geq 1 \, ,
\end{equation}
where $\rho_h^{(j,k)}$ is the halo field density for channel $j$ at pixel $k$ and ${\boldsymbol{\theta}}(x)$ is the Heaviside function. The threshold $T_c^{(j)}$ is determined from the empirical CDF of  $\rho_{\text{h}}^{(j,k)}$, in such a way that 
\begin{equation}
    P(\rho_{\text{h}}^{(j,k)}> T_c^{(j)}) = 0.05 \,. 
\end{equation}
In other words, the threshold eliminates pixels that contain low-density values, and only retains the highest 5\% of the log-density distribution in each channel. Note that for a given channel $j$, the pixels used for the computation are taken not only from the channel $j$, but also from the immediate neighboring channels, $j + 1$ and $j-1$. For the boundary channels (i.e., $j=1$ and $j=4$), we include only the next or previous channel, respectively. Now we define a pixel- and channel-specific mask $m_w^{(i,k)}$ by incorporating these weights:
\begin{equation}
    m_w^{i,k} = 
\begin{cases} 
    w^i & \text{if } \rho_{\text{h}}^{(i,k)} > T_c^{(i)} \\
    1  & \text{if } \rho_{\text{h}}^{(i,k)} \leq T_c^{(i)} \, .
\end{cases}
\end{equation}
We normalize the mask such that $\displaystyle \sum_{i,k} m_w^{i,k}= 64$, or each training mask of shape $4\times64^3$ adds to 1 on average, since each simulation is made of 64 sub-volumes.

\begin{figure*}
    \centering
    \includegraphics[width=1.0\linewidth]{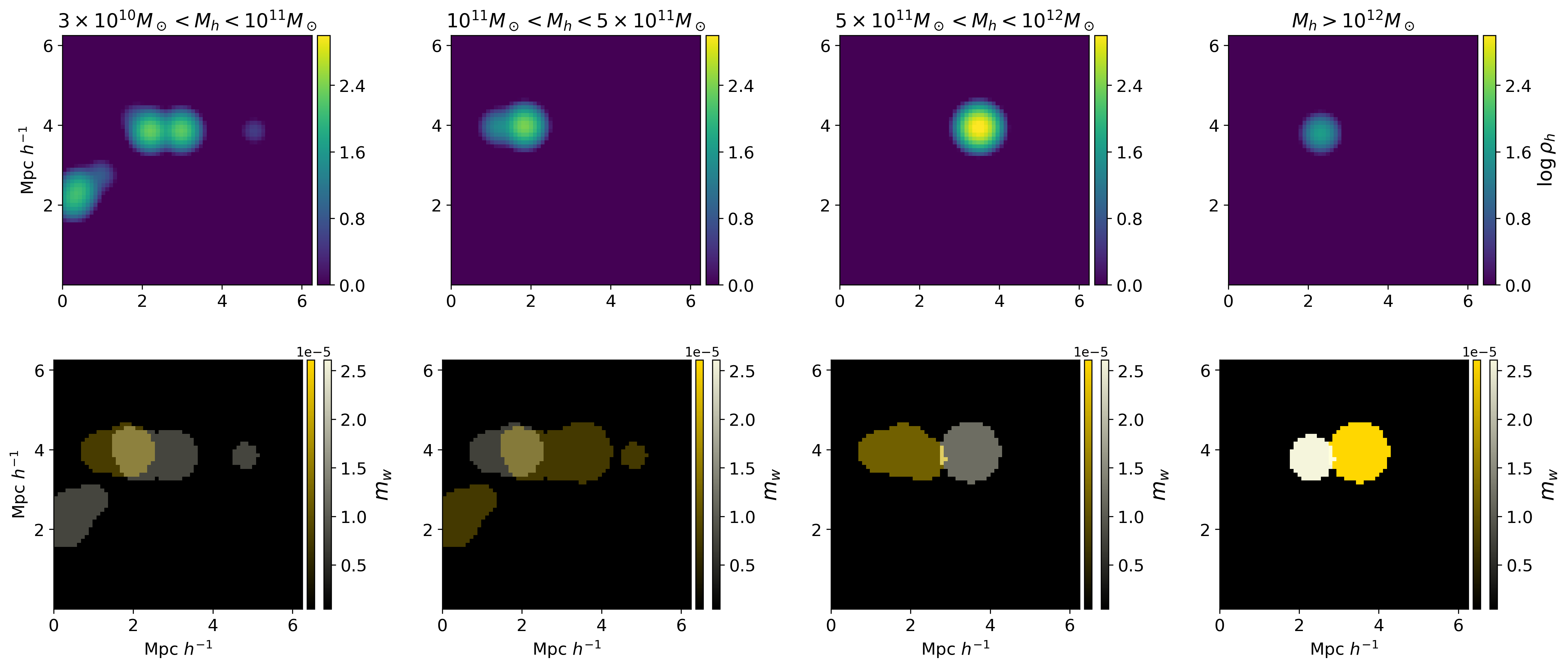}
    \caption{Illustration of the halo density channels masking prescription. The first row of 4 images shows a 2D slice of the training halo maps $\rho_\text{h}^{(j)}$ for each halo mass channel $j$. The second row shows the corresponding 2D weighted mask maps, $m_w^{j,k}$. In each masking map, the contribution from the halo $j$ is shown in beige, and the contribution from the neighboring channel's halo location is shown in golden.}
    \label{fig:unet_halo_MASK}
\end{figure*}

An illustration of the construction of $m_w^{(i,k)}$ is shown in figure~\ref{fig:unet_halo_MASK}. By construction, pixels corresponding to halo sites for channel $j$ (beige) have high mask values; the mask also has large values for adjacent channel's halo positions (golden), in order to enforce the absence of halos there in the particular channel under consideration, since  for these regions, the target voxel value corresponds to the small background values.

The masked loss function is then written as:
\begin{equation} \label{eqn:lossfn}
    \mathcal{L} = \displaystyle\sum_{j=1}^4 \left(\alpha L_2^j\odot m_w^j\beta^j + \left(1-\alpha \right)L^j_q
     \right) + \lambda_w L_1^{\text{reg}}
\end{equation} 
with $\alpha,\beta^{j=1, \dots, 4},\lambda_w$ being hyper-parameters of the model. The loss incorporates both the mean squared error $L_2^j$ and the quantile loss $L_q^j$  between the pixel-level target density, $\rho_{\text{h}}^{(j,k)}$, and its predicted value, $\rho_{\text{pred}}^{(j,k)}$, weighed by the mask for the $L_2$ loss, with the direct product running over pixel values $k$:
\begin{equation}
\begin{split}
L_{q}^{j} &=
  \frac{1}{N_\text{mesh}}\sum_{k=1}^{N_\text{mesh}}
  \max\!\Bigl(
      q\left(\rho_{\text{h}}^{(j,k)}-\rho_{\text{pred}}^{(j,k)}\right),
      (q-1)\left(\rho_{\text{h}}^{(j,k)}-\rho_{\text{pred}}^{(j,k)}\right)
  \Bigr),\\
L_{2}^j &=
  \frac{1}{N_\text{mesh}}\sum_{k=1}^{N_\text{mesh}}
  \left(\rho_{\text{pred}}^{(j,k)}-\rho_{\text{h}}^{(j,k)}\right)^{2}.
\end{split}
\end{equation}
Here $q \in (0, 1)$ is the chosen quantile.
The weights $\beta^j$ set the relative scaling of the contribution of a particular channel to the loss. 
To further deal with sparsity, we add the $L_1^{\text{reg}}$ regularization term, defined as:
\begin{equation}
    L_1^{\text{reg}} = \sum_{i=1} \left| w^i_{\boldsymbol{\theta}} \right|
\end{equation}
where $\left| w^i_{\boldsymbol{\theta}} \right|$ are the neural network weights, and $i$ runs over the network's parameters.
\vspace{\baselineskip}

\subsection{\LODI: HI intensity maps with latent overlap diffusion} \label{sec:halo_to_HI}

\subsubsection{Diffusion model architecture}

Diffusion models are a class of generative models, proposed by~\citet{sohl-dickstein_deep_2015} and \citet{ho_denoising_2020} that start from a Gaussian random field and denoise it in steps to produce a data point sampled from the underlying data distribution. A diffusion model has two parts: in the forward noising process, Gaussian noise is progressively added to a data point to convert it into white noise. This forward process can be written as,
\begin{equation}\label{eq:forward_diffusion}
    q\left( \textbf{x}_t|\textbf{x}_0 \right) = \mathcal{N} \left( \alpha_t\textbf{x}_0,\sigma^2_t \mathcal{\textbf{I}} \right)
\end{equation}
where $\textbf{x}_0$ is the original data point and $\textbf{x}_t$ is the noisy version of $\textbf{x}_0$ at timestep $t$,  where $t \in \{1, 2, \dots, \text{T}\}$ and $T$ is the total number of noising steps. {Here, $q\left( \textbf{x}_t|\textbf{x}_0 \right)$ denotes the probability distribution of the noisy sample $\textbf{x}_t$, given an original data point $\textbf{x}_0$, The notation $\mathcal{N} \left( {\mathbf{\mu}}, \mathbf{\Sigma}\right)$ indicates a multivariate Gaussian distribution with mean $\mathbf{\mu}$ and covariance matrix $\mathbf{\Sigma}$. } The noise is added according to a schedule defined by $\alpha_t$ and $\sigma^2_t$. In the reverse process, also known as the generation process, the model learns to conditionally denoise this added noise in $T$ steps with a general form:
\begin{equation}\label{eq:denoising}
    p_{\boldsymbol{\theta}}\left( \textbf{x}_{t-1}|\textbf{x}_t \right) = \mathcal{N} \left( \textbf{x}_{t-1}; \bm{\mu}_{\boldsymbol{\theta}}\left( \textbf{x}_t,t \right),\mathbf{\Sigma}_{\boldsymbol{\theta}}\left( \textbf{x}_t,t \right)\right)
\end{equation}
\vspace{\baselineskip}
Here, {$p_{\boldsymbol{\theta}}( \textbf{x}_{t-1}|\textbf{x}_t )$ denotes the learned reverse (generative) distribution parameterized by the denoising neural network parameters, ${\boldsymbol{\theta}}$. The distribution can also be parametrized as a multivariate Gaussian with covariance $\mathbf{\Sigma}_{\boldsymbol{\theta}}$ and mean $\bm{\mu}_{\boldsymbol{\theta}}$, obtained by the same denoising neural network that was trained to predict $\textbf{x}_{t-1}$ given $\textbf{x}_t$. This formulation holds generally for both denoising diffusion probabilistic models (DDPMs) and variational diffusion models (VDMs), the latter being the framework adopted in this work. }  The loss function for this task minimizes an objective of the form:
\begin{align}
\mathcal{L}_T(\mathbf{x}) &=  \mathbb{E}_{\epsilon, t} \left[ \left( f_{\boldsymbol{\theta}}(\sigma_t,\alpha_t\right) \|\bm{\epsilon} - \hat{\bm{\epsilon}}_{\boldsymbol{\theta}}(\mathbf{x}_t; t)\|_2^2 \right] \, ,
\end{align}
where $\bm{\epsilon}$ and $\hat{\bm{\epsilon}}$ are the added and predicted noise, respectively. {Here, $\mathbb{E}_{\epsilon, t}[\cdot]$ is the expectation value with respect to the Gaussian distribution of the noise $\epsilon$ and timestep $t$ sampled from the diffusion process, and $\|\cdot \|_2^2$ denotes the squared $l_2$ norm. The term $f_{\boldsymbol{\theta}}(\sigma_t,\alpha_t)$ is a weighing function which depends on the noising variance and scaling factors. It weighs the relative contribution of a given denoising step at time $t$ to the total loss, and in our case its explicit form is given in equation \eqref{lossdiff}. }
\vspace{\baselineskip}

\begin{figure*}
    \centering
    \includegraphics[width=0.6\linewidth]{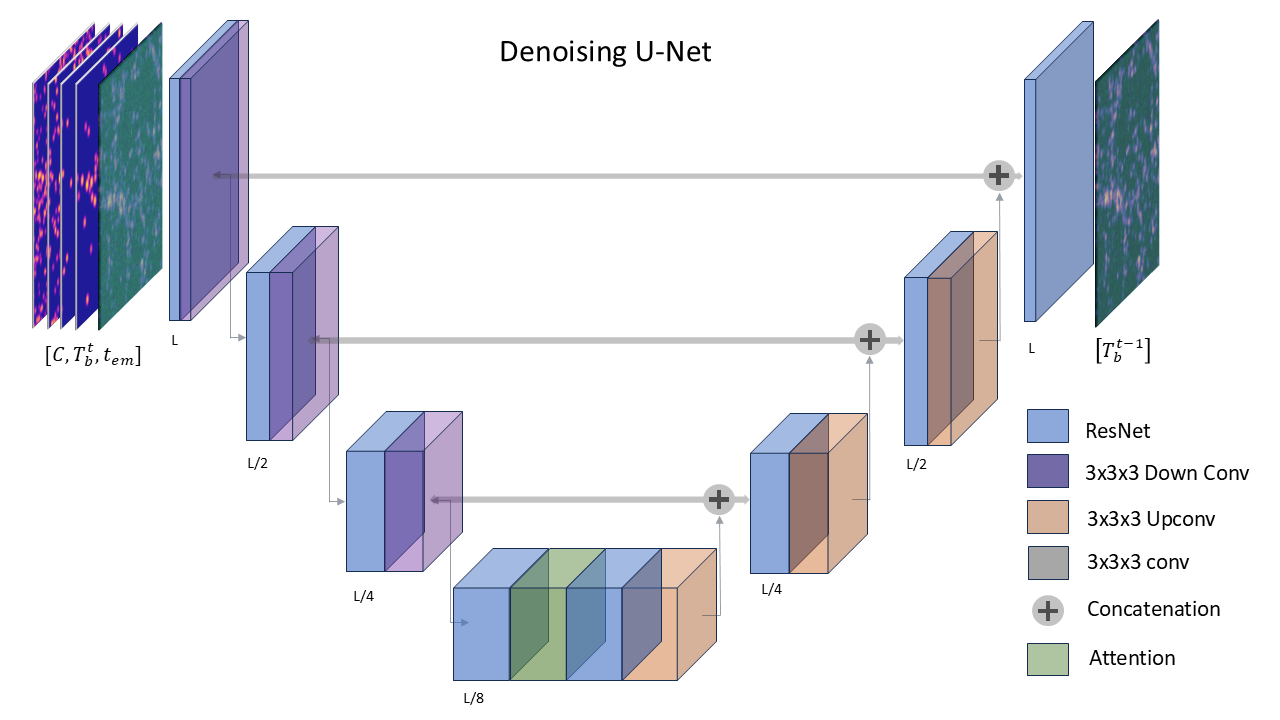}
    \caption{The architecture of the denoising model: a 3D U-net with residual network blocks and an attention layer at the bottleneck. The input to the network is the noisy temperature map at time step $t$, the time step embedding and the conditional halo maps. The output is a less noisy version of the temperature map.}
    \label{fig:DenoiseUnet}
\end{figure*}

To inpaint the HI brightness temperature, $T_b$, onto the halo distribution obtained in the previous step, we use the variational diffusion model (VDM) of~\citet{kingma_variational_2023}. The neural network architecture is modified to work with 3-dimensional simulation boxes and follows a similar architecture as the one provided in~\citet{ono_debiasing_2024}:. 2D grouped convolutions are replaced with their 3D counterpart, and we add an extra convolutional layer to the residual blocks of the network. 
We set up the denoising architecture as a U-Net with residual networks and an attention block at the bottleneck, as illustrated in Fig.~\ref{fig:DenoiseUnet}. This is because it has been shown that attention blocks at the bottleneck facilitate faster convergence and achieve the same quality of cross correlation results with fewer training steps~\citep{ono_debiasing_2024}.

\subsubsection{Building the 21cm temperature training map}
\label{sec:2.2.2}
To produce the 21 cm temperature map for training, we proceed as follows: {we obtain the neutral hydrogen fraction $x_{\text{HI}, i}$} and the gas particle mass $m_{g, i} \,  [M_\odot]$, for each gas particle $i$ in the hydrodynamic simulation, from which we {first compute the HI mass per particle, $M_{\text{HI}, i}$, as: } 
{
\begin{equation}
    M_{\text{HI}, i} = m_{g,i}x_{\text{HI}, i}X \, ,
\end{equation}
}
{where $X$ is the hydrogen mass fraction, which in this CAMELS simulations is set to $0.76$}. In addition, since the simulations (as detailed below) do not include self-shielding effects for star-forming particles, we manually set $x_{\text{HI}, i} = 1$ for particles with star formation rate larger than zero. {The reason for having a separate star formation rate parameter on top of the star particles is that in the underlying multiphase ISM model, self-shielded gas is represented as a gas element until stochastically converted to a star particle, where the conversion probability depends on the star formation timescale~\citep{Springel_2003}.} This simple and effective prescription, also employed in the post-processing of the CAMELS simulations~\citep{villaescusa-navarro_camels_2022}, assumes that all the star-forming particles are fully self-shielded against ionizing radiation. As shown in ~\citet{villaescusa-navarro_ingredients_2018}, this assumption results in a good match to observational data sets like abundance of HI and column density distribution of Damped Lyman-$\alpha$ systems.

{We then use the same CIC procedure described above to interpolate $M_{\text{HI}, i} $ to a continuous mass field, $M_\text{HI}$, after which we divide each voxel by its volume to obtain the density field $\rho_{\text{HI}}$}. The 21 cm brightness temperature in the post-reionization universe is given to a good approximation by the following expression in~\citet{villaescusa-navarro_ingredients_2018}:
\begin{equation}\label{tb}
    T_b(\textbf{x}) = 189h\left(\frac{H_0(1+z)^2}{H(z)}\right)\frac{\rho_{\text{HI}}(\textbf{x})}{\rho_c} ~[\text{mK}]\, , 
\end{equation}
{where $H_0$ is the Hubble-Lema\^itre constant today, $G$ is the gravitational constant, $\rho_{\text{HI}}(\textbf{x})$ is the interpolated neutral hydrogen density at a given cell location $\textbf{x}$}, $z$ is the redshift, $H(z)$ is the Hubble function at redshift $z$, $T_{\rm b}$ is expressed in units of mK, and $H_0 = 100h$ km/s/Mpc. The quantity {$\rho_c = \frac{3H_0^2}{8\pi G}$ denotes the critical density of the Universe today, i.e. the density required for a spatially flat Universe.}

We will work in the plane-parallel approximation and assume the redshift of the box as a constant, given the fact that redshift evolution within the simulated volume would require a realistic set of lightcones mock and this is beyond the scope of the paper.
Once the maps are created, we smooth them with a Gaussian kernel of fixed radius $R = 0.2 ~\text{Mpc $h$}^{-1}$. {This choice balances two considerations: on the one hand, smoothing helps to stabilize training by reducing sharp fluctuations and noise in the target fields; on the other hand, excessive smoothing would wash out the small-scale structures that are physically important for our task. Given that the voxel resolution is $\approx 0.1 \mpc$, we chose a slightly larger smoothing radius that facilitates improved training of the network while preserving most of the relevant small-scale information. }

\subsubsection{Loss function}
\label{diffusion_explain}
When training \LODI, we first sample ${\bf T}_b$ map from the training set and then apply to it the forward diffusion process of Eq.~\eqref{eq:forward_diffusion}, setting ${\bf x}_0 = {\bf T}_b$, ${\bf x}_t = {\bf T}_b^t$, $\alpha_t^2 = $ sigmoid$(-\gamma_{\boldsymbol{\eta}}(t) )$ and $\sigma_t^2 = $ sigmoid$(\gamma_{\boldsymbol{\eta}}(t) )$, where the noising schedule $\gamma_{\boldsymbol{\eta}}(t) = b+wt$ with learnable parameters $\boldsymbol{\eta} \;=\; \{\,w,\,b\,\}$. For our work, we use $T = 50$.

During learning, the network predicts the noise that was added at step $t$ by minimizing the diffusion loss:
\begin{align}
\label{lossdiff}
    \mathcal{L}_T(\textbf{T}_b) = \frac{T}{2} \, \mathbb{E}_{\epsilon \sim \mathcal{N}(\mathbf{0},\mathbf{I}), \, t \sim U\{1,T\}} \bigg[ &
    \big(\exp\big(\gamma_{\boldsymbol{\eta}}(t-1) - \gamma_{\boldsymbol{\eta}}(t)\big) - 1\big) \notag \\ 
    & \times \big\|\boldsymbol{\epsilon} - \boldsymbol{\hat{{\epsilon}}_{\boldsymbol{\theta}}}(\textbf{T}_b^t; t) \big\|_2^2
    \bigg].
\end{align}
 where $\mathcal{L}_T(\textbf{T}_b)$ is the diffusion loss, which measures the error in noise prediction; {where the expectation is taken over the noise distribution $\boldsymbol{\epsilon} \sim \mathcal{N}\!\bigl(\mathbf{0},\,\mathbf{I}\bigr)$ and $t \sim U[1,T]$, denoting the uniform discrete distribution.} The noise predicted by the de-noising UNet is $\boldsymbol{\hat{\epsilon}_{\boldsymbol{\theta}}}(\textbf{T}_b^t; t)$. Thus, in the forward process, noise is added until we obtain pure Gaussian noise at timestep $T$, when $\textbf{T}_b^T \sim \mathcal{N}(\mathbf{0},\mathbf{I})$. During the generation process, we sample from a standard normal and denoise it in $T$ steps using the learned distribution $p_{\boldsymbol{\theta}}(\textbf{T}_b^{t-1}|\textbf{T}_b^{t},\boldsymbol{\rho}_\text{h},t)$ to generate a sample $\textbf{T}_b^0 \sim p(\textbf{T}_b|\boldsymbol{\rho}_\text{h})$. Sampling from this learned distribution is equivalent to performing ancestral sampling~\citep{kingma_variational_2023}, which in this case simplifies to:
\begin{equation}
\label{ancestral}
    \textbf{T}_b^{s} = \frac{\alpha_s}{\alpha_t}\left(\textbf{T}_b^{t}-\sigma_t c \boldsymbol{\hat{\epsilon}_{\boldsymbol{\theta}}}(\textbf{T}_b^t; t)\right) + \sqrt{(1-\alpha_s^2)c}\boldsymbol{\epsilon}
\end{equation}
where $c = \exp \left(\gamma_\eta(s) - \gamma_\eta(t)-1 \right)$, $\boldsymbol{\epsilon}\sim \mathcal{N}(\boldsymbol{0},\mathbf{I})$ and $0<s<t<T$.

\subsection{Datasets, training and validation}
\label{dataset}
We use the CAMELS dataset~\citep{villaescusa-navarro_camels_2022}, which is a suite of hydrodynamical simulations created by varying cosmological and astrophysical parameters, along with different initial seeds. This is especially useful to test the generalization performance of the model trained on 3D cosmological dark matter density and brightness temperature maps. {Each simulation box has a comoving side of length $25h^{-1}$ Mpc and contains $256^3$ gas cells and $256^3$ dark matter particles. CAMELS is based on the AREPO~\citep{weinberger_arepo_2020} and GIZMO~\citep{Hopkins_2015_GIZMO} codes and run the same subgrid physics and feedback effects of IllustrisTNG and SIMBA simulations respectively. The hydrodynamics is evolved with the moving-mesh code AREPO  which combines a TreePM gravity solver with a finite-volume Godunov scheme on an unstructured Voronoi mesh. Gas cells above a density threshold are treated with the  multiphase star-formation model of \citet{Springel_2003}. Stellar winds and black-hole seeding/growth with AGN feedback
follow the IllustrisTNG prescriptions. The particles are evolved from a redshift of $z=127$. } 

For this work, we train on the CV set of IllustrisTNG, which contains $256^3$ gas and dark matter particles (which are later subdivided into $256^3$ voxels), all run with the cosmological and astrophysical parameters set to: $\Omega_m=0.3,\sigma_8 = 0.8,A_\text{SN1}=A_\text{SN2}=A_\text{AGN1}=A_\text{AGN2}=1.0$, {Here, $\Omega_m$ denotes the present-day matter density parameter and $\sigma_8$ is the root-mean-square amplitude of linear matter fluctuations on scales of $8h^{-1}$ Mpc. $A_\text{SN1}, A_\text{SN2}$ are the supernova feedback process parameters of the model and $A_\text{AGN1}, A_\text{AGN2}$ are the active galactic nuclei feedback process parameters, with values of 1.0 corresponding to their fiducial calibration in IllustrisTNG.} There are 27 different initial seeds for each of the 27 simulations.

\subsubsection{Training of \HALO}
To train \HALO, we use one single simulation from the CV set and test on other simulations in the same set, with different initial seeds, but same simulation parameters. Both dark matter and halo maps are divided into $64^3$ sub-volumes, rotated, and augmented to get rotationally-equivariant training. 

We input the $64^3$ voxels $\boldsymbol{
\rho}_\text{DM}$ field in batches of 16 and apply the loss function {in equation}~\eqref{eqn:lossfn} to the predicted output, using the weighted mask in {equation}~\eqref{eqn:mask}. The optimizer used for training is {RMSprop (Root Mean Square Propagation), an adaptive gradient-based optimization algorithm that scales the learning rate for each parameter by a running average of its recent squared gradients, a standard technique in deep learning architectures.} The learning rate schedule is set as follows: the first 10 epochs are run with a learning rate of $2\times10^{-4}$; after that, we use Cosine Annealing Warm Restarts with the maximum and minimum values set to $10^{-4}$ and $3\times10^{-5}$, respectively, with a repeat cycle of 50 epochs. {The cosine annealing warm restarts changes the learning rate between the maximum and minimum value, with a cosine decay, while repeating this cycle every 50 epochs in our case.} We run the model for 260 epochs, and store the model's weights at the point in which the validation loss is lowest. We subdivide the data into training and validation set, with a $75\%$ and $25\%$ split. 
{In equation~\eqref{eqn:lossfn},} the coefficient $\lambda_w$ of the regularization term $L_1^{\text{reg}}$ is set to $10^{-6}$. For the results in the paper, we use $\alpha = 0.6,q=0.7$ and $\beta^j=[1,2,5,8]$. The above hyper-parameters have been chosen to achieve the best predictions, after extensive testing. 

\subsubsection{Training of \LODI}

We train \LODI\ on 5 simulations of the CV set from CAMELS, each with a different initial seed. We process the 5 simulations as explained above, and divide them into $32^3$ voxel sub-volumes.
To enforce rotational equivariance, we augment each simulation cube with all symmetry-equivalent flips and 90° rotations. Training on more simulations is necessary to account for cosmic variance in the large scales comparable to the box size and also to train on more halos of large masses, especially $M_\text{halo}>10^{12}M_\odot$.  This expands the raw training set substantially and would correspondingly increase training time. We therefore apply a targeted sub-sampling strategy: because HI emission is concentrated around dark matter halos --particularly massive ones-- and thus sparse, we preferentially draw sub-volumes that contain objects in the higher–mass halo channels. We define important regions here as regions where $\log \rho_h\geq 10$, since this density value is about $10^3$ times smaller than the minimum density value in our dataset, prior to gaussian smoothing. Values smaller than that are small enough that they do not contribute to any meaningful HI content. To maintain diversity we also include $50\%$ of sub-volumes centered on the more numerous lower-mass halos, the first 2 mass bins in this case. This balanced selection reduces the total training data sets from $\approx$ 410 000 to $\approx$ 250 000, almost halving the  training time without degrading the representativeness of the data, while increases the relative importance of the larger mass halos as it is seen more frequently by the neural network.  We finally create a training and validation split of $75\%$ and $25\%$, respectively.

\subsection{Generation over larger volumes with latent overlap}

It is not possible to input the entire $25^3$  $\left(\text{Mpc}/{h}\right)^3$ simulation box, corresponding to $256^3$ voxels to our network, because of memory constraints. Such data cube would require several terabytes of GPU memory to train. Instead, we divide the domain into 512 smaller $32^3$ voxels sub-volumes, with the view of generating for each its HI density field with the trained the diffusion model, and then combine them at the end to generate the full volume. 

{\subsubsection{Latent overlap method}}
The naive solution would be to generate the brightness temperature field in each $32^3$ sub-volume, then tile them to create the target, $256^3$ volume -- we call this approach `tiling'. However, tiling introduces artificial, periodic discontinuities at the boundaries of each $32^3$ voxel cube, arising because the brightness temperature field $\textbf{T}_b$ is sampled from the conditional distribution  $\textbf{T}_b\sim p(\textbf{T}_b|\boldsymbol{\rho}_\text{h})$, which depends solely on the $32^3$ halo field $\boldsymbol{\rho}_h$, with no reference to adjacent boxes. 

\begin{figure*}
   \centering
   \includegraphics[width=1.0\linewidth]{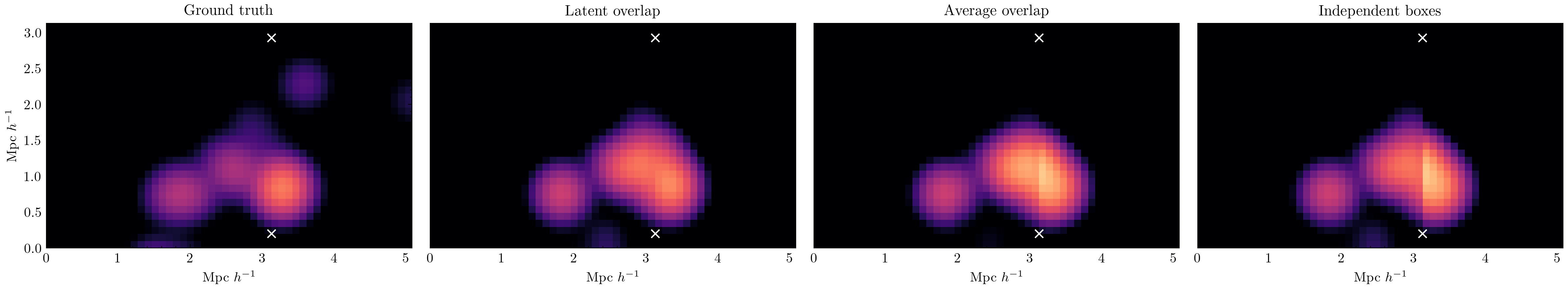}
   \caption{A comparison of different overlapping methods are shown, where the line of discontinuity is vertical and passes through the 2 marked crosses in the images. The left image is the ground truth. The latent overlap method is shown in the left center, compared with two other cases. The center right is the case where for the overlap region, we take the average pixel values of the two boxes and the right image shows the other case when we generate 2 separate boxes and just concatenate them, while keeping the the overlapping pixel region from the first box. The images shown are averaged in the third axis over a distance of about 0.9 $\text{Mpc} h^{-1}$}
   \label{fig:cont_proof_comp}
\end{figure*}

To address this, we adapt the inpainting method of ~\citet{lugmayr_repaint_2022}, which in our work we call `latent overlap' approach. 
The first box, $B_1$, is generated by starting with a Gaussian noise $\boldsymbol{\epsilon_1}$, to obtain a sample $\textbf{T}^0_{B_1}$. While generating this box in $T=50$ steps, a set of $T$ intermediate latent maps are produced using Eq.~\eqref{ancestral}: $[\textbf{T}^1_{B_1},\textbf{T}^2_{B_1},\dots,\textbf{T}^T_{B_1}]$. These intermediate latent images can be interpreted as discrete snapshots of the evolving $\textbf{T}_b$  distribution under the diffusion process, forming a trajectory from random noise to the final sampled field. 

When we next consider an adjacent box, $B_2$, we want to enforce that its generation process follow a `similar' trajectory as $B_1$. For this, we partially overlap boxes $B_2$ onto $B_1$ such that
\[
\text{Vol}(B_1 \cap B_2) = V_{\text{overlap}}
\]
where $V_{\text{overlap}}$ is the overlapping volume and is defined as the fraction of one box (specified below). {For our work, the overlap volume or fraction is specified in section \hyperref[21cm_overlap]{2.4.2}.} Now, at every $t$-th denoising step of $B_2$, we replace $\textbf{T}^t_{B_2}$ with the previously-generated $\textbf{T}^t_{B_1}$ in the region $V_{\text{overlap}}$, thus obtaining $\textbf{T}^{t*}_{B_2}$, which is identical to $\textbf{T}^t_{B_1}$ in the overlap volume. This new $\textbf{T}^{t*}_{B2}$ is then denoised using the denoising U-net to obtain $\textbf{T}^{t-1}_{B_2}$. 
Mathematically, the updated state of \( B_2 \) at the \(t\)-th denoising step is given by:
\[
\textbf{T}^{t*}_{B_2} = M_{\text{overlap}} \textbf{T}^t_{B_2} + (1 - M_{\text{overlap}}) \textbf{T}^t_{B_1} \, ,
\]
where \( M_{\text{overlap}} \) is the overlap mask defined as:
\[
M_{\text{overlap}}(\textbf{x}) =
\begin{cases} 
1, & \text{if } \textbf{x} \in V_{\text{overlap}} \\
0, & \text{otherwise}
\end{cases}
\]
and
\[
\textbf{T}^t_{B_1} = \alpha_t \textbf{T}^0_{B_1} + \sigma_t \boldsymbol{\epsilon}, \quad \boldsymbol{\epsilon} \sim \mathcal{N}(\textbf{0}, \textbf{I}) \, ,
\]
where $\alpha_i$ and $\sigma_i$ are the noise schedule parameters and are learnt by the diffusion model as explained in section~\ref{diffusion_explain}, and $\textbf{T}^0_{B_1}$ was produced by the diffusion model in a previous backward pass.

\begin{figure*}
    \centering
    \includegraphics[width=0.8\linewidth]{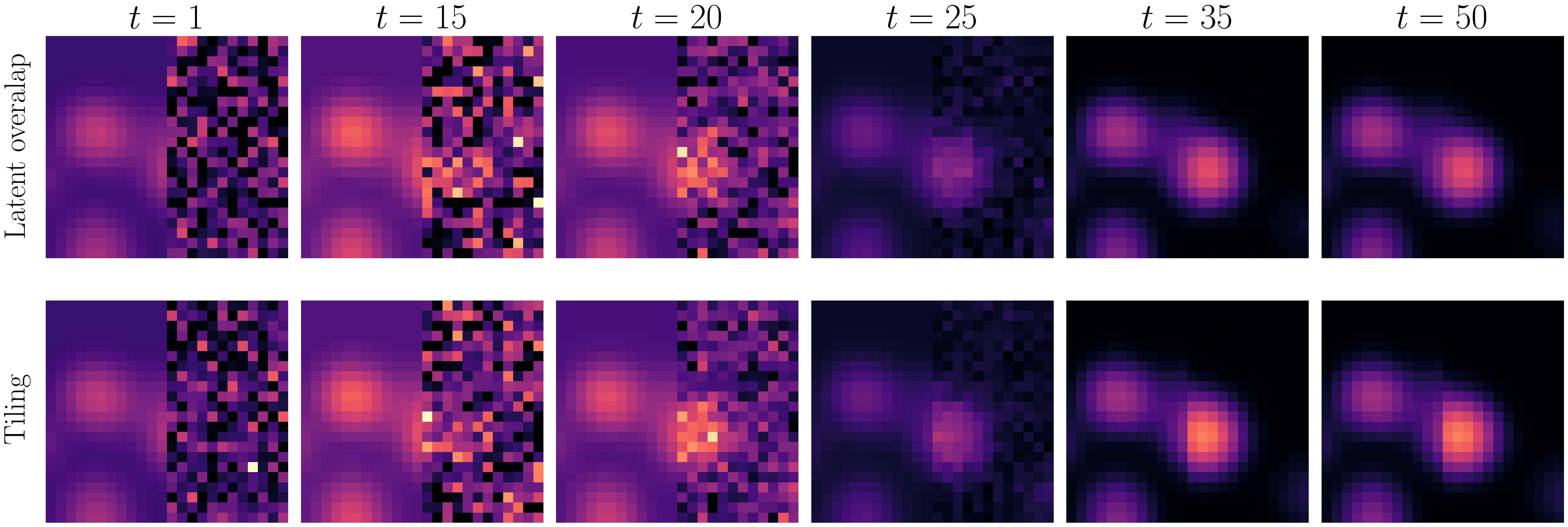}
    \caption{Illustration of \LODI, using the latent overlap method: output of the diffusion process at different steps in the denoising process (top row, left to right) when using the latent overlap method on a 2D illustrative slice. At each time step, the left half of the field has already been generated. The bottom row shows the same time steps but with simple tiling and no latent overlap: the discontinuity at the boundary is apparent.}
    \label{fig:algo_proof}
\end{figure*}

Performing this latent overlap step ensures that the entire image $\textbf{T}^{i*}_{B_2}$ is denoised harmoniously, thereby eliminating boundary artifacts, while producing a non-overlapping region that fits seamlessly with the overlapping region. We carry out this procedure $T$ times, and at each step $B_2$ is pushed closer to the distribution of $B_1$ by forcing it to align with the overlapping region of $B_1$ while moving along the trajectory of $\textbf{T}^{i}_{B_1}$. This latent overlap method between boxes $B_1$ and $B_2$ is shown in Algorithm~\autoref{alg:sample}, while figure \ref{fig:cont_proof_comp} provides a comparison of the results obtained with latent overlap against simple tiling of independent boxes. In principle, as $T$ is increased, the alignment should improve. Figure \ref{fig:algo_proof} shows the diffusion process implementing this algorithm on a 2D patch of overlapping boxes. In this case, the left box has already been produced and does not change throughout the entire process\footnote{Note that left half of the image does not change during the process, despite the misleading impression given by the colorbar boundaries being adjusted from left to right to accommodate the different dynamic range of pixel values.}. We call this approach \LODI, for Latent Overlap Diffusion for Intensity mapping.

\begin{algorithm}
\caption{Latent overlap method}
\label{alg:sample}
\begin{algorithmic}[1]
\State $\textbf{T}_{B_2}^T \sim \mathcal{N}(\textbf{0}, \textbf{I})$
\State $\boldsymbol{\epsilon_{B_1}}\sim \mathcal{N}(\textbf{0}, \textbf{I})$
\State Calculate $M_{\text{overlap}}$
\For{$i = T-1, \dots, 0$}
    \State $ s(i)=(i)$ \textbf{and} $t(i)=i+1$
    \State $\textbf{T}^t_{B_1} = \alpha^t \textbf{T}^0_{B_1} + \sigma^t \boldsymbol{\epsilon_{B_1}}$
    \State $\textbf{T}^{t*}_{B_2} = M_{\text{overlap}} \textbf{T}^t_{B_2} + (1 - M_{\text{overlap}}) \textbf{T}^t_{B_1}$
    \State $\boldsymbol{\epsilon_2}\sim \mathcal{N}(\textbf{0}, \textbf{I})$
    \State $\textbf{T}_{B_2}^{s} = \frac{\alpha_s}{\alpha_t}\left(\textbf{T}_{B_2}^{t*}-\sigma_t c \boldsymbol{\hat{\epsilon}_{\boldsymbol{\theta}}}(\textbf{T}_{B_2}^{t*}; t)\right) + \sqrt{(1-\alpha_s^2)c}\boldsymbol{\epsilon_2}$
    
\EndFor
\State \textbf{return} $T_{B_2}^0$
\end{algorithmic}
\end{algorithm}
\vspace{\baselineskip}

{\subsubsection{Patching 21 cm fields}}
\label{21cm_overlap}

To generate the entire $25^3$ $\left(\text{Mpc}/h\right)^3$ volume, we need to define the sequence in which the sub-volumes $B_i$ cubes are processed. Ideally, we aim for an algorithm that enables parallelization of the process, for faster generation, while using the largest possible number of starting sub-volumes (i.e., the ones that are generated conditionally on the learnt diffusion process, but without reference to adjacent regions), because these are the ones that are truly in-distribution w.r.t. to the diffusion generative process.

\begin{figure*}
    \centering
    \includegraphics[width=0.6\linewidth]{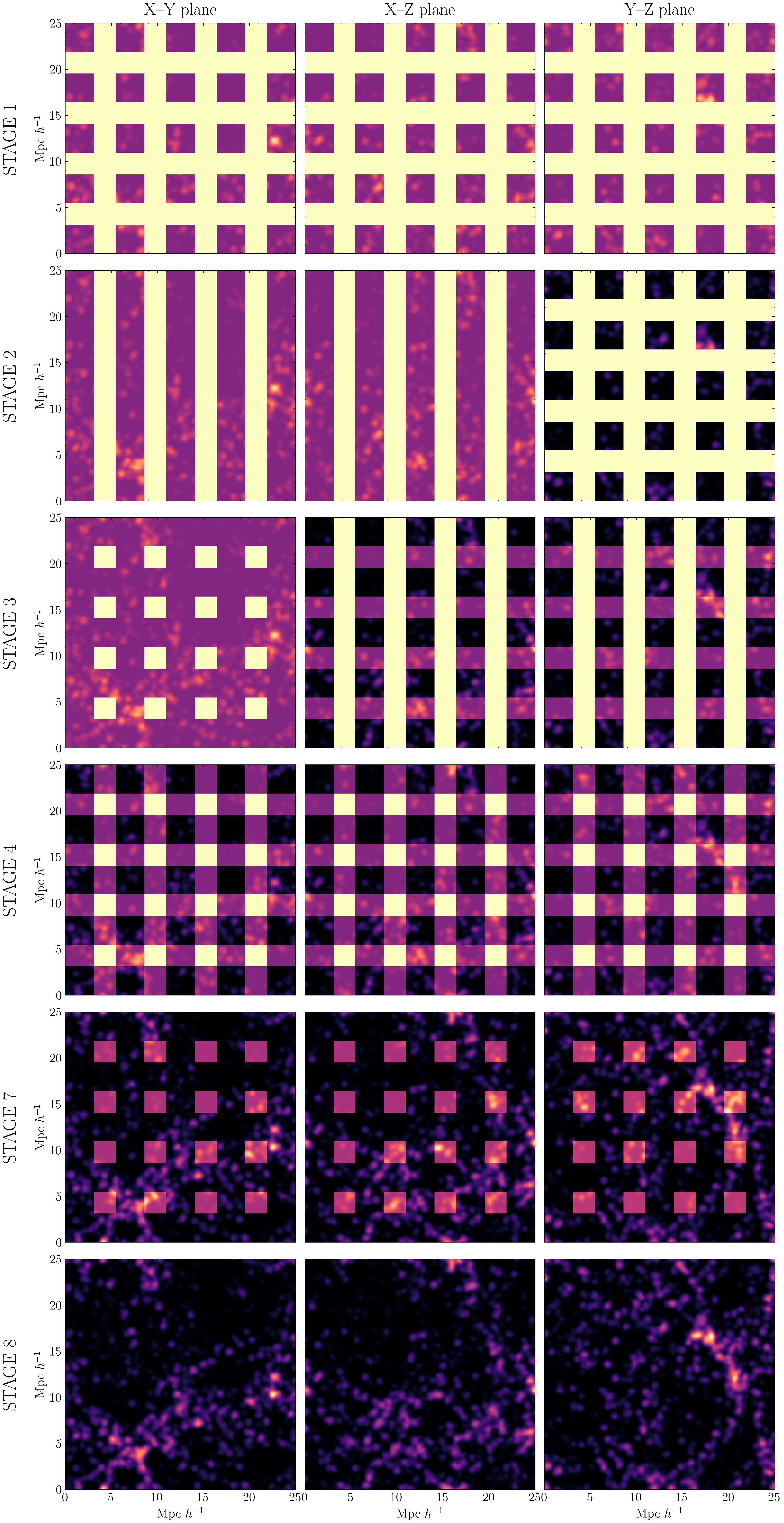}
    \caption{Assembly of the full volume cube of voxel size $256^3$ from $9 \times 9 \times 9$ sub-volumes using the latent overlap method, for a simulation from the validation set. Each row corresponds to a different stage (top to bottom) of the latent generation process and each column displays one of the three orthogonal 3D cartesian planes of the simulation volume ({the maps are shown by projecting along the third  spatial axis across the full $25 \mpc$ extent of the simulation box}). Although the figure displays six rows, the complete process consists of eight sequential steps. The fifth row represents the net result of three consecutive sub-steps, which have been aggregated for clarity.     \label{fig:overlapping_scheme}}
\end{figure*}

To execute this, we create a mesh of 125 sub-volumes $B_i$ of size $32^3$ voxels,  equally spaced from each other, arranged in a 3D grid (Stage 1, top panels in Fig.~\ref{fig:overlapping_scheme}, showing from left to right the three orthogonal planes, $X-Y$, $X-Z$, $Y-Z$; maps are averaged along the third axis). In Stage 2, 3 and 4, we generate nearest-neighbor boxes along each of the 3 cartesian direction using the latent generation method with an overlap of $4\times32\times32$ voxels. This greatly reduces the out-of-distribution error, and helps with parallelizing the generation process since most of the boxes can be generated independently of each other. Consequently, the configuration in Stage 4 exhibits an alternating pattern where every other $B_i$ is present along each axis. In each of these 3 stages, 100 new $B_i$s are generated per axis, amounting to 300 new generations in total.

In the subsequent Stages $(5,6~\text{and}~ 7)$, additional sub-volumes are generated adjacent to the ones generated in stages 2,3 and 4 along the cartesian axes, overlapping with the original sub-volume with the same overlap fraction of $4\times32\times32$. This procedure is performed sequentially for the 3 cartesian directions, requiring 80 cubes per axis (240 in total). Finally, the remaining 64 sub-volumes are generated at stage 8, each of which overlaps with the neighboring 6 voxels with the same overlap volume of $4\times32\times32$ to complete the configuration, resulting in a final assembly of 729 $B_i$s arranged in a $9 \times 9\times 9$ grid resulting in a total volume of $25$ $\left(\text{Mpc}/h\right)^3$. This boosts the speed of generation by a factor of $\approx$6, compared to generating sub-volumes sequentially. In principle we could further parallelize the generation process by combining the different stages into one, but this is limited by the available GPU memory.

\begin{figure*}
    \centering
    \includegraphics[width=0.8\linewidth]{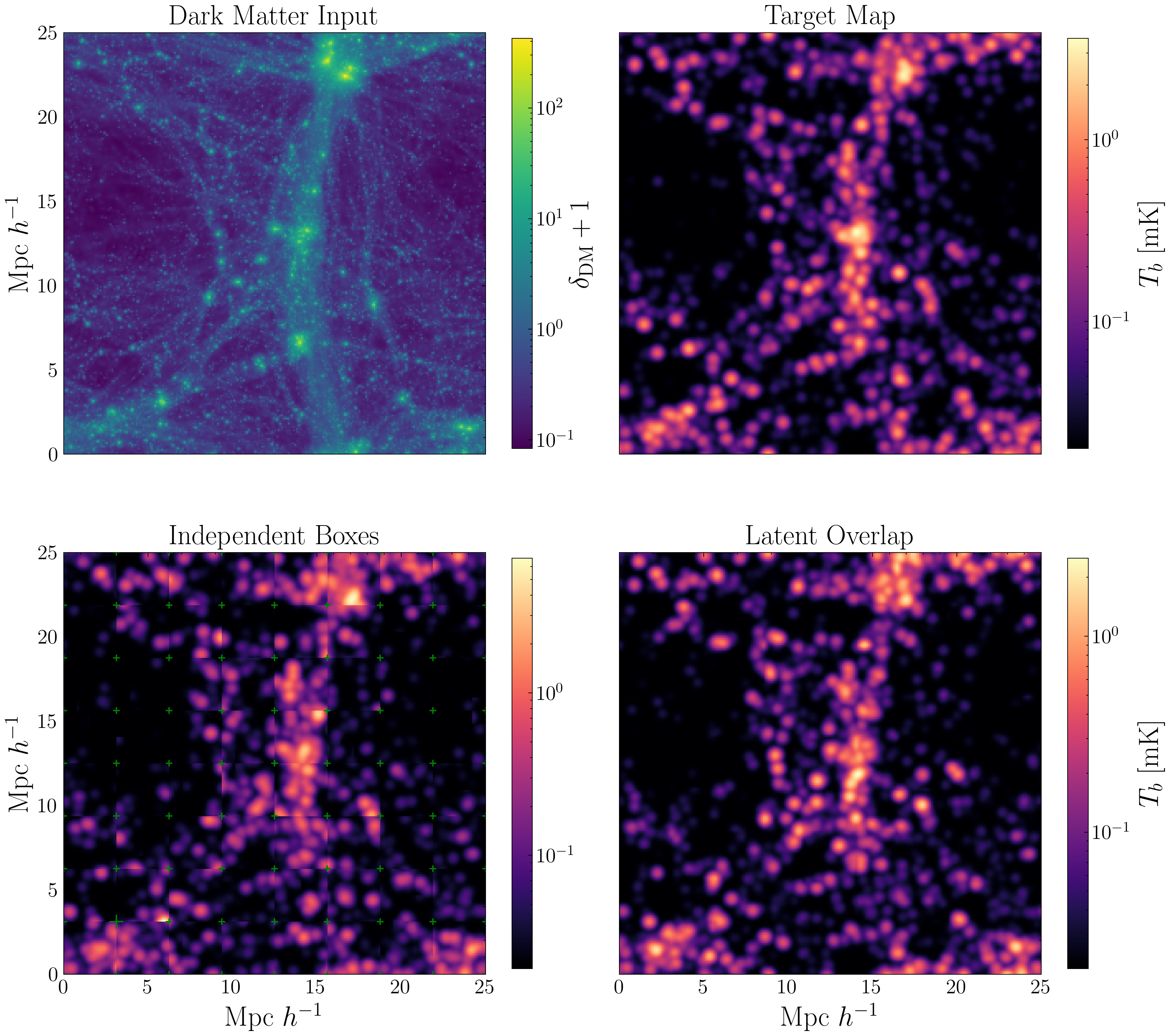}
    \caption{Top left: The dark matter density map, not seen during training, used as input for the generative pipeline. Top right: the ground truth (target) HI brightness temperature map from the simulation. Bottom left: The generated output using the Tiling method, with green markers highlighting the sub-volumes boundaries where discontinuities arise. Bottom right: brightness temperature map generated using the latent overlap method of \LODI, which significantly reduces the discontinuities seen in the tiling approach. The brightness temperature map is conditioned on the output provided by \HALO. {All the maps are shown by projecting along the third  spatial axis across the full $25 \mpc$ extent of the simulation box}. }
    \label{fig:overall_output}
\end{figure*}

The overall output of our generative pipeline is shown in Fig.~\ref{fig:overall_output}. We start with an unseen  dark matter density field, shown in the top left panel, run it through \HALO\ (whose intermediate halo mass density maps are shown in the second column of figure~\ref{fig:Halo_finding}), and then produce the brightness temperature map with \LODI\ (bottom right panel of Fig.~\ref{fig:overall_output}). Using the naive Tiling approach (bottom left panel of Fig.~\ref{fig:overall_output}), the discontinuity between training sub-volumes is evident (corners of boundaries are indicated by green crosses for better visualization).

\section{Results}
\label{results}
In this section, we demonstrate the performance of the entire generative pipeline, producing HI intensity maps from previously unseen dark matter simulation maps. We validate our results using the CV set of the IllustrisTNG simulations at $z=0$, with the same parameters for the simulations as the training maps, but with different initial seeds. We postpone the study of generalization capabilities to other cosmologies to a future, dedicated work. 

\subsection{Performance of  \HALO }
\begin{figure*}
    \centering
    \includegraphics[width=0.75\linewidth]{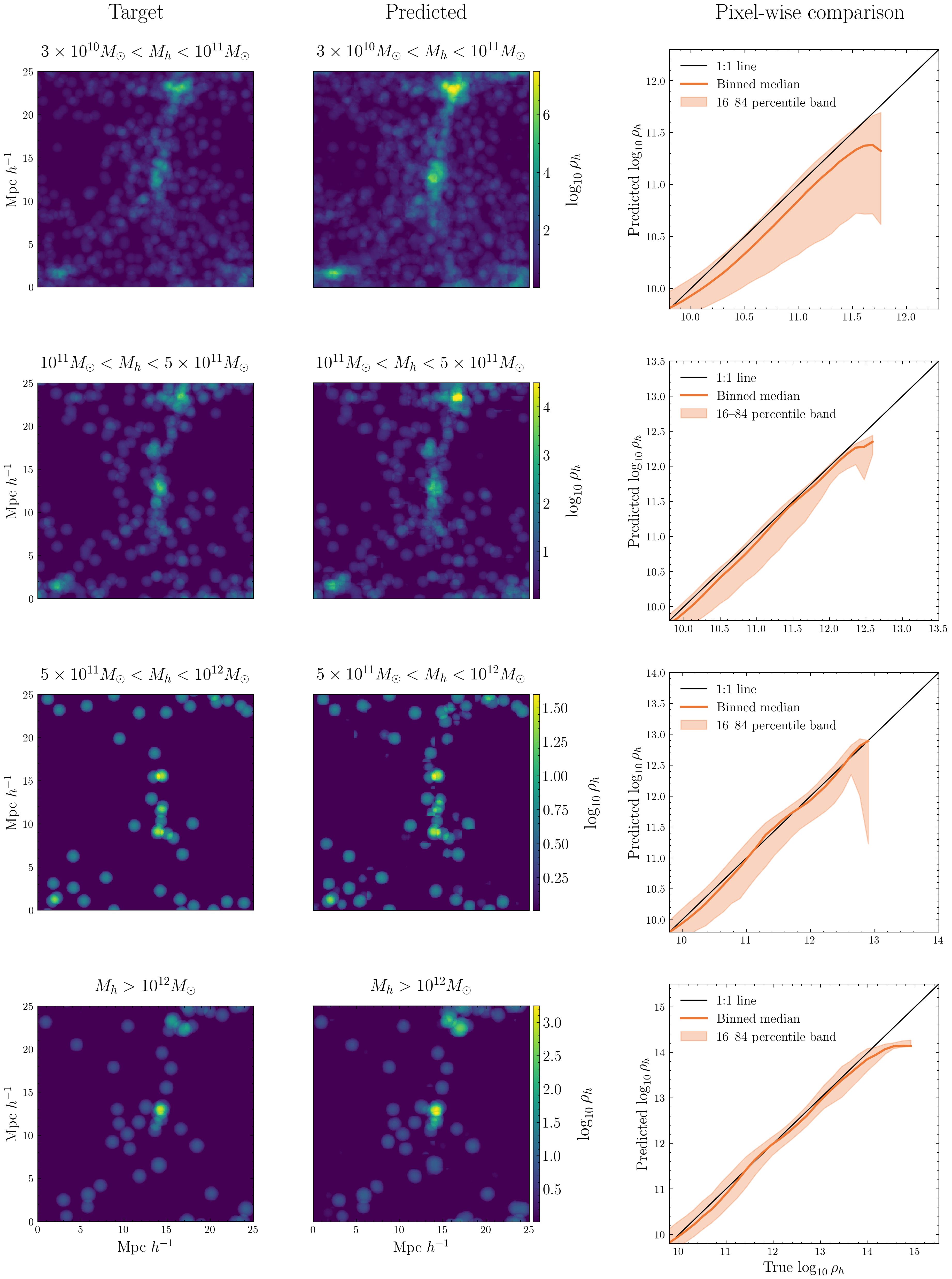}
    \caption{
    Output of \HALO\ from a previously unseen dark matter simulation.  The rows show the comparison between the target (left) and predicted (centre) halo mass density maps for each of the four mass channels (top to bottom). {The halo maps are visualized by projecting along the third spatial axis across the entire $25 \mpc$ of the simulation.} The right column gives a pixel-wise comparison between the distribution of log-densities for the target and predicted map, with the mean and 1$\sigma$ band computed from {the $256^3$ voxels}.}
    \label{fig:Halo_finding}
\end{figure*}

The output of \HALO\  is shown in figure~\ref{fig:Halo_finding}, split in the four mass channels (top to bottom), with the input dark matter map shown on the top left of figure~\ref{fig:overall_output}. We reiterate that this input map was unseen during training, as it was generated with a different seed (but same cosmological and astrophysical parameters as the training map). 

Visually, the predicted pixel values for the halo mass density (left column) look very close to the target maps (center column), except for the smallest mass range (first row). This can be more precisely quantified by comparing the distribution of log-density values between the target and predicted $25^3$ $\left(\text{Mpc}/\text{h}\right)^3$ box (right column). While we observe a consistent under-prediction in the mean value of the smallest halo mass range (albeit still within the $16-84$ percentile band), the agreement for the other mass ranges is excellent. We trace back the relatively poorer performance for the smallest mass halo density to our definition of masked loss: as discussed above, a lower weight is given to pixels of the smallest halos, since they are the most abundant ones. Therefore, we can expect that the loss function puts greater attention to faithfully reconstructing the density of larger mass halos. We plot the median instead of the mean since in the post-processing of the output halo maps, we set pixel values below a threshold of $\log\rho_h = 9.5$ to zero, which in the pixel comparison plot skews the mean value of the pixel distribution. The median values are agnostic to these outlier near zero values. The reason for setting this cutoff is because such low predicted values are not possible given the training data and lie below the lower density limit.

\subsection{Performance of  \LODI}

We define the dimensionless brightness temperature field 
by normalizing each spatial point \( T_b(\mathbf{x}) \) by its global mean 
\(\langle T_b \rangle\):
\begin{equation}
  \hat{T}_b(\mathbf{x}) \;=\; \frac{T_b(\mathbf{x})}{\langle T_b \rangle} \, .
\end{equation}
The dimensionless power spectrum is accordingly defined as: 
\begin{equation}
  P_{\hat{T}_b}(k) 
  \;=\; 
  \Bigl\langle\, \hat{T}_b(\mathbf{k}) \cdot \hat{T}_b^{*}(\mathbf{k}) \Bigr\rangle,
\end{equation}
where \(\hat{T}_b(\mathbf{k})\) is the Fourier transform of 
\(\hat{T}_b(\mathbf{x})\). This dimensionless formulation allows for 
a direct comparison of small-scale fluctuations in the brightness 
temperature field across different simulations.

To assess the performance of \LODI, we analyze the 21cm power spectrum produced by our pipeline 
and its residual,
\[
  \Delta P_{\hat{T}_b}(k) 
  \;=\; 
  P_{\hat{T}_b}^{\mathrm{gen}}(k) 
  \;-\; 
  P_{\hat{T}_b}^{\mathrm{true}}(k).
\]
where $P_{\hat{T}_b}^{\mathrm{gen}}(k)$  and $P_{\hat{T}_b}^{\mathrm{true}}(k)$ denote 
the \LODI-generated and ground truth (from the simulation) 21 cm dimensionless power spectrum, respectively. Since the diffusion model was trained on Gaussian smoothed maps, the comparison plots and power spectra calculations are also done by Gaussian smoothing the ground truth maps with the same filter radius of $R = 0.2 ~\text{Mpc $h$}^{-1}$. {Note that when testing our model, we compare against maps that include HI from all the halos across the full halo-HI mass range. This allows us to verify
that the specific halo-mass threshold used during training does not
significantly affect the recovered 21cm signal.}

\begin{figure*}
    \centering
    \includegraphics[width=0.6\linewidth]{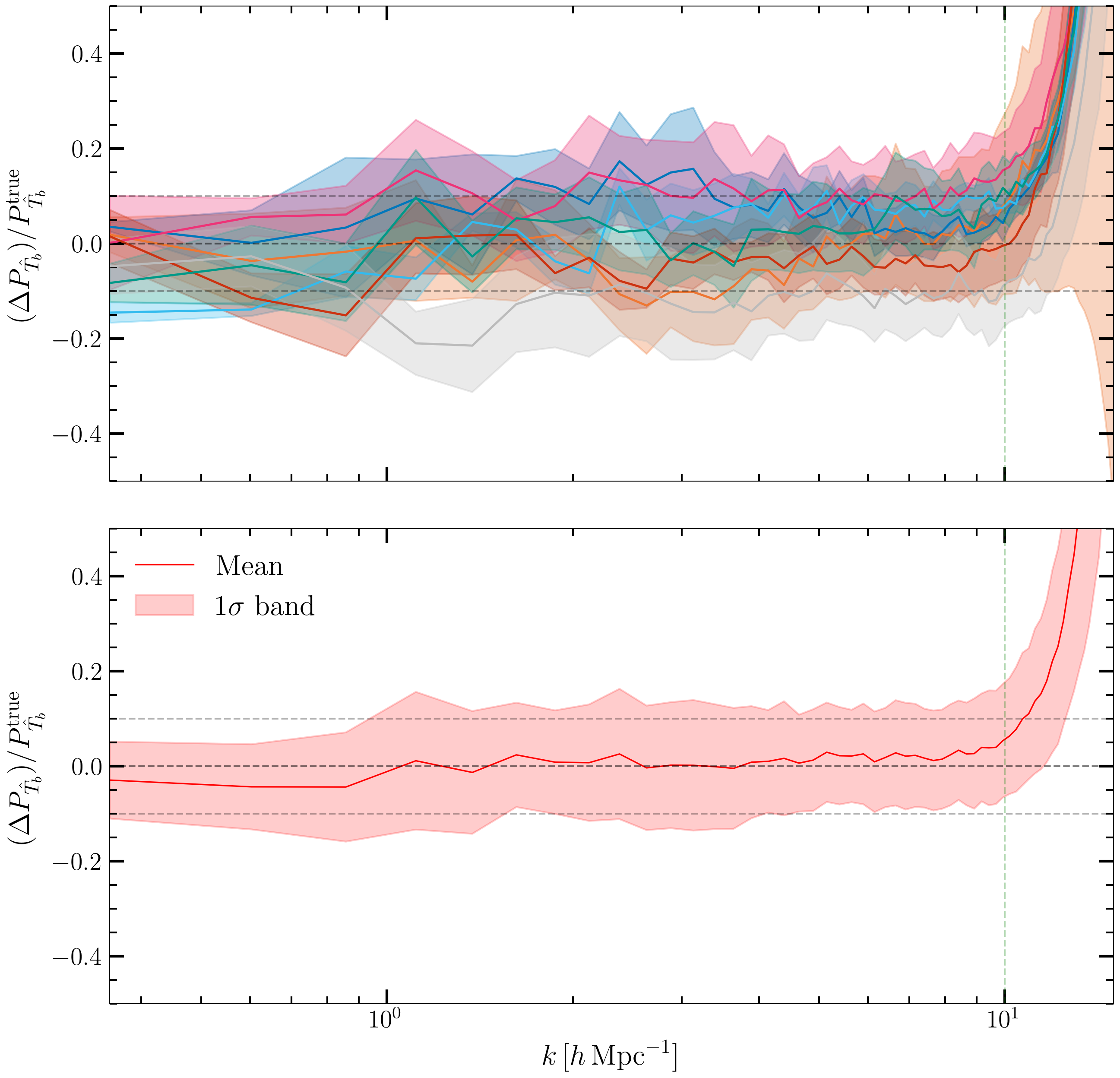}
    \caption{Top: mean (lines) and standard deviation (shaded bands) of the residual $T_b$ power spectrum for 7 dark matter simulations (each colour a different seed), with 5 diffusion realizations each. The bottom plot shows the result when combining all the realizations.
    }
    \label{fig:pk_dif_many}
\end{figure*}

Figure~\ref{fig:pk_dif_many} illustrates the end-to-end performance by showing the residual power spectrum obtained by applying the whole pipeline on 7 previously unseen dark matter density fields (i.e., with different initial seeds). For each of the 7 dark matter fields, we first produce the halo maps for each using \HALO, and then generate 5 realizations of brightness temperature maps.

\begin{figure*}
    \centering
    \includegraphics[width=0.8\linewidth]{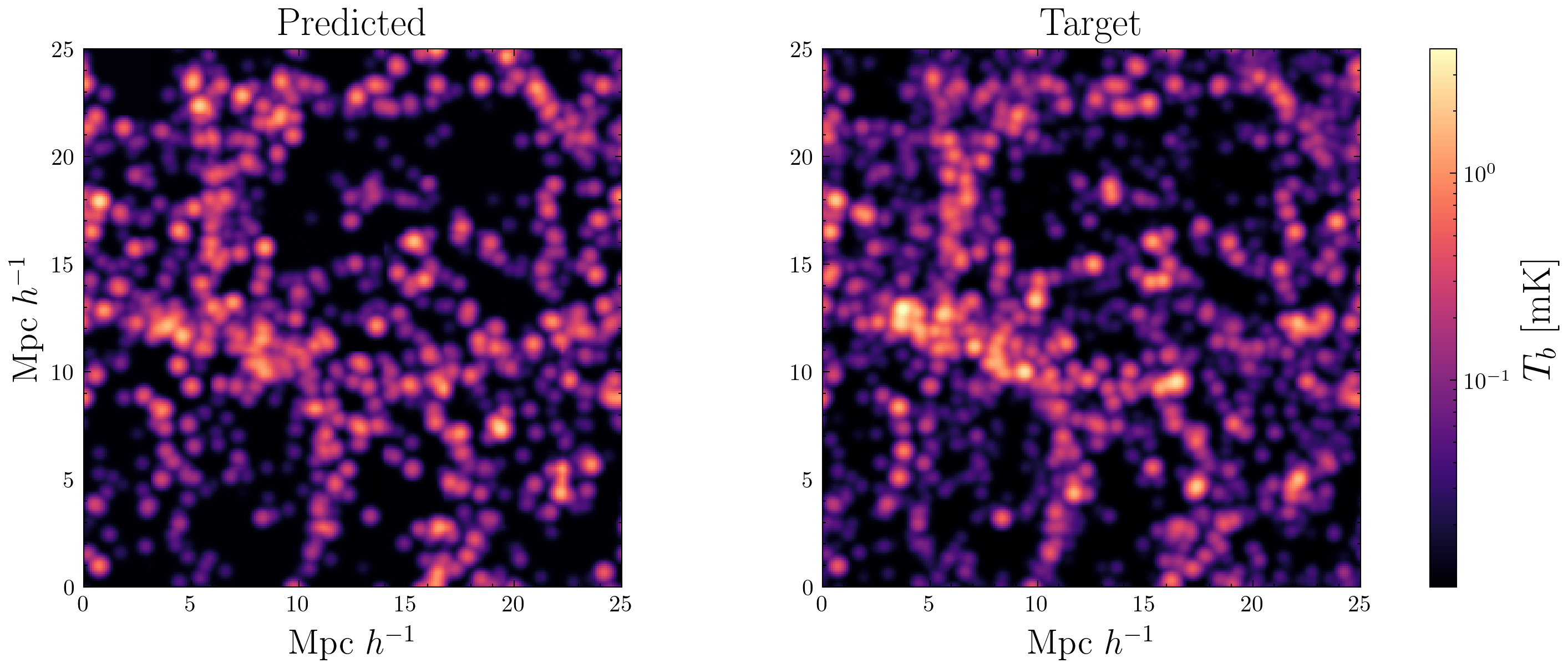}
    \hspace*{-1cm}\includegraphics[width=0.6\linewidth]{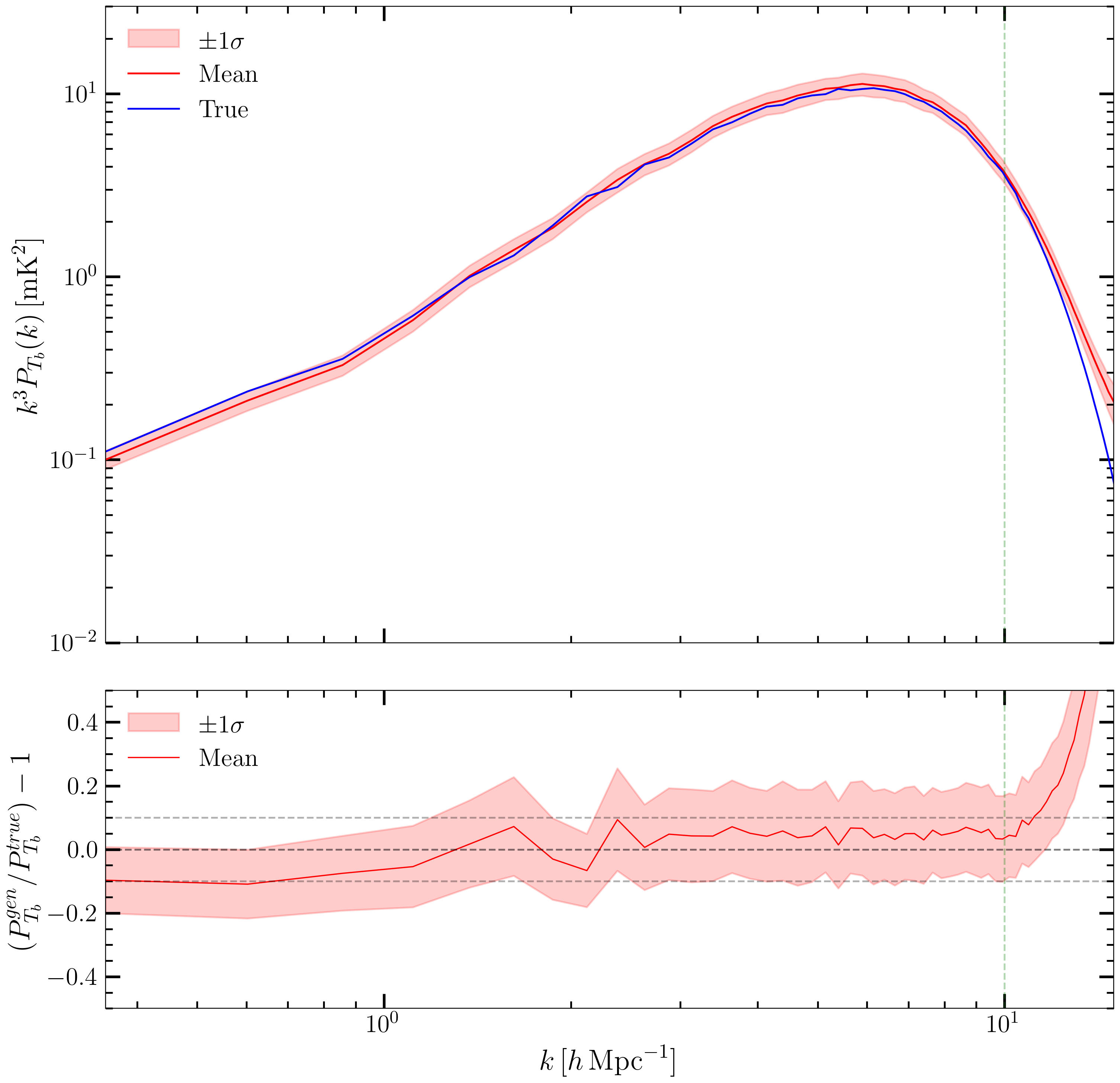}
    \caption{Generation of brightness temperature maps with \LODI:  one sample of  diffusion-generated map (top left) compared to the unseen ground truth (top right) { visualized by averaging over the third spatial axis, along the entire $25 \mpc$ length of the simulation box}. The bottom panel compares the mean power spectrum and its standard deviation obtained from 20 diffusion realizations, each produced with a different seed for the diffusion model, showing good agreement well into the non-linear regime, up to scales $k \simeq 10\text{\it h} ~\text{Mpc}^{-1}$, also visible in the residual spectra shown in the bottom panel. The maps are smoothed with a Gaussian kernel of radius of $0.2~\text{\it h}~\text{Mpc}^{-1}$.}
    \label{fig:pk_same_many}
\end{figure*}

In Figure~\ref{fig:pk_same_many}, we present an example $T_b$ map generated using the full pipeline, alongside its corresponding ground-truth map, unseen before during training, each averaged along the 
third axis. We also plot the dimensional power spectrum $P_{T_b}$, computed from 20 \LODI\ realizations conditioned on the same halo maps produced by \HALO. Visually, the generated and true maps show considerable matching, and the power spectra remain in good agreement up to $k \simeq 10h ~\text{Mpc}^{-1}$, as can also been in the residual values shown in the bottom panel. This result showcases the intrinsic variance of the generative model \LODI, as seen in the spread of the power spectra residuals. {At high wavenumbers the predicted power spectrum exhibits excess power relative to the truth. This arises because the generated 21 cm maps display more compact radial profiles than the true maps, which are comparatively more extended and diffuse. Since \LODI\ is conditioned on halo maps only, the network tends to concentrate 21 cm emission within halo regions, producing higher central intensities and a steeper radial decline. This enhanced contrast amplifies small-scale variance and leads to the observed excess power at higher wave modes.}

To generate these results, \HALO~was trained for about 21 hours on a single A100 GPU. The training of \LODI\ took  600,000 iterations, using a batch size of 16. The total training time was around 30 hours, using a single A100 GPU. 

When generating, the total time taken by the pipeline to produce one temperature map for the whole $256^3$ volume starting from the dark matter field is 106 seconds (16 seconds for \HALO, 90 seconds for \HALO).

\section{Conclusions}
\label{conclusions}
In this work, we presented a two-steps generative pipeline to generate and inpaint large, 3 dimensional 21 cm intensity maps, starting from dark matter only simulations. Although this work applied the method to  25$^3$ $\left(\text{Mpc}/\text{\it h}\right)^3$ simulations, it can be extended to inpaint simulations of any size, provided the resolution is the same as the training set simulation. In the first step, the \HALO\  algorithm uses a U-Net to produce high-fidelity halo maps, subdivided in mass channels (bins); in the second step, \LODI\ uses a conditional variational diffusion model coupled with a latent overlap method to paint the 21 cm signal on top of the halo field. We also developed a method to parallelize the generation of sub-volumes of intensity maps coherently into a larger volume simulation, without having to modify the diffusion model. Thus, this method is agnostic to the type of diffusion model used. A more powerful and general method is being investigated and will be demonstrated in a dedicated future work.

To create the halo maps, we use an attention ResUnet along with a weighted masking technique, to enforce halo separation based on halo mass. In this work, we create 4 halo maps from a single dark matter density field, which we then use as a conditional in our diffusion model -- a variational diffusion model that learns to denoise a Gaussian field, conditioned on the 4 halo maps, to produce the 21 cm intensity maps at redshift $z=0$. 

The pipeline produces statistically unbiased estimates of the 21 cm power spectrum, and achieves a $\leq 10\%$ accuracy on average, up to $k \simeq 10\,\it{h} ~\text{Mpc}^{-1}$, as can be seen in figure~\ref{fig:pk_dif_many}. We test the model on multiple unseen simulations, with different initial seeds, showcasing the robustness of the model to capture and reproduce features well into the non-linear small-scale regime.
\vspace{\baselineskip}

Although the model performs well on simulations that share the same redshift, cosmology, and astrophysical prescriptions, it is essential to test its ability to generalize across various cosmological parameters and redshift. Such a robust model would enable parameter inference and simulation-based, likelihood-free analyses, and would allow us to build full past-light-cone mocks, that could be compared with data, potentially increasing the information retrieval many-folds compared with simpler (and lossy) summary statistics, like the power spectrum. In addition, forthcoming large-volume HI surveys will enable high-signal-to-noise cross-correlations between 21 cm intensity maps and galaxy catalogs ~\citep{Cunnington:2022}, an approach that mitigates foreground systematics and can tighten constraints on both cosmology and galaxy astrophysics. Our pipeline can therefore be extended to produce realistic HI mocks for these surveys, providing a test-bed for cross-correlation analyses and improved joint constraints on cosmological and astrophysical parameters. We also foresee to extend the pipeline presented here by coupling $\HALO$ and $\LODI$ with \texttt{JERALD}, a Lagrangian deep learning method that produces high-resolution dark matter, stellar mass and neutral hydrogen maps from lower-resolution approximate N-body simulations~\citep{JERALD}, thereby further accelerating the end-to-end analysis. 

Furthermore, it will be interesting to investigate adapting the diffusion model to cope with realistic foregrounds and instrumental nuisances (like the role of the beam, wedge reconstruction, etc.), both in terms of cross-correlation with galaxies and auto-correlation of the 21 cm brightness temperature. Finally, from the fundamental physics perspective, the excellent performance of \LODI\ even at very small scales makes it an ideal tool to explore non-standard dark matter scenarios, in which the dark matter could have thermal velocities or interactions with baryons and radiation.

\section*{Acknowledgments}
The authors thank Danijel Skočaj for helpful suggestions that inspired the methods adopted, as well as Bruce Bassett, Daniela  Breitman, Kosio Karchev, David Prelogovič, Mauro Rigo, Giulio Scelfo, Francisco Villaescusa-Navarro and Gabrijela Zaharijas for discussions. 
SM thanks the Flatiron CCA and University of Nova Gorica for hospitality.

SM is supported by the National Recovery and Resilience Plan (PNRR), Dottorati Green/Innovazione under DM 351 and also acknowledges support from the SISSA-Flatiron Exchange Programme. 
RT acknowledges co-funding from Next Generation EU, in the context of the National Recovery and Resilience Plan, Investment PE1 Project FAIR ``Future Artificial Intelligence Research''. This resource was co-financed by the Next Generation EU [DM 1555 del 11.10.22]. RT and MV are partially supported by the Fondazione ICSC, Spoke 3 ``Astrophysics and Cosmos Observations'', Piano Nazionale di Ripresa e Resilienza Project ID CN00000013 ``Italian Research Center on High-Performance Computing, Big Data and Quantum Computing'' funded by MUR Missione 4 Componente 2 Investimento 1.4: Potenziamento strutture di ricerca e creazione di ``campioni nazionali di R\&S (M4C2-19)'' - Next Generation EU (NGEU).  Part of the simulations were postprocessed on the Ulysses supercomputer at SISSA.
MV and RT are also partially supported by the INFN INDARK grant.

No generative AI was used in the writing of this article.

\section*{Data Availability}

The full source code, implemented in Python using PyTorch, along with usage instructions and notebook scripts is publicly available at \url{https://github.com/satvik-97/LODI}. All data used to produce the results in this paper are available upon request.



\bibliographystyle{mnras}
\bibliography{references,Diffusion_21cm}


\bsp	
\label{lastpage}
\end{document}